\documentclass[reprint, amsmath, amssymb,aps,pra,reprint,superscriptaddress]{revtex4-2}
\usepackage{graphicx, color}  
\usepackage{dcolumn}   
\usepackage{bm}        
\usepackage{amssymb}   
\usepackage{amsmath}   
\usepackage{footmisc}
\usepackage{enumerate}
\usepackage{comment}
\usepackage{multirow} 
\usepackage{tabularx}
\usepackage{booktabs}
\usepackage{placeins}

\usepackage{hyperref}
\usepackage{siunitx}
\usepackage{multirow} 

\newcommand{\smallMatrix}[1]{\begin{pmatrix}#1\end{pmatrix}}
\newcommand{\RN}[1]{%
  \textup{\uppercase\expandafter{\romannumeral#1}}%
}

\begin{document}

\preprint{APS/123-QED}

\title{Study of the optical losses as a function of beam position on the mirrors in a 285 m suspended Fabry-Perot cavity}

\author{Y. Zhao}
\affiliation{Universit\'e Paris Cit\'e, CNRS, Astroparticule et Cosmologie, F-75013 Paris, France}
\author{M. Vardaro}
\affiliation{Maastricht University, 6200 MD Maastricht, Netherlands}
\affiliation{Nikhef, 1098 XG Amsterdam, Netherlands}
\author{E. Capocasa}
\thanks{eleonora.capocasa@apc.in2p3.fr}
\affiliation{Universit\'e Paris Cit\'e, CNRS, Astroparticule et Cosmologie, F-75013 Paris, France}
\author{J. Ding}
\affiliation{Universit\'e Paris Cit\'e, CNRS, Astroparticule et Cosmologie, F-75013 Paris, France}
\affiliation{Corps des Mines, Mines Paris, Universit\'e PSL, France}
\author{Y. Guo}
\affiliation{Maastricht University, 6200 MD Maastricht, Netherlands}
\affiliation{Nikhef, 1098 XG Amsterdam, Netherlands}
\author{M.  Lequime}
\affiliation{Aix Marseille Univ, CNRS, Centrale Med, Institut Fresnel, Marseille, France}
\author{M. Barsuglia}
\affiliation{Universit\'e Paris Cit\'e, CNRS, Astroparticule et Cosmologie, F-75013 Paris, France}

\date{\today}

\begin{abstract}

Reducing optical losses is crucial for reducing quantum noise in gravitational-wave detectors. In fact, with equal input power, a lower level of round trip losses in the arm cavities allows to store more power. Moreover losses are the main source of degradation of the squeezed vacuum, which along with power increasing, is the most effective strategy to reduce quantum noise. Frequency dependent squeezing obtained via a filter cavity is currently used to reduce quantum noise in the whole detector bandwidth. Such filter cavities are required to have high finesse in order to produce the optimal squeezing angle rotation and the presence of losses is particularly detrimental for the squeezed beam, as it does multiple round trip within the cavity. Characterising such losses is crucial to assess the quantum noise reduction achievable. In this paper we present an \textit{in-situ} measurement of the optical losses, done for different positions of the beam on the mirrors of the Virgo filter cavity. We implemented an automatic system to map the losses with respect to the beam position on the mirrors finding that optical losses depend clearly on the beam hitting position on input mirror, varying from 42\,ppm to 87\,ppm, while they are much more uniform when we scan the end mirror (53\,ppm to 61\,ppm). We repeated the measurements on several days, finding a statistical error smaller than $\pm 4$\,ppm. The lowest measured losses are not much different with respect to those estimated from individual mirror characterisation performed before the installation (30.3 - 39.3\,ppm). This means that no major loss mechanism has been neglected in the estimation presented here. The larger discrepancy found for some beam positions is likely to be due to contamination.
In addition to a thorough characterisation of the losses, the methodology described in this paper allowed to find an optimal cavity axis position for which the cavity round trip losses are among the lowest ever measured. This work can contribute to achieve the very challenging losses goals for the optical cavities of the future gravitational-wave detectors, such Einstein Telescope and Cosmic Explorer.

\end{abstract}

\maketitle

\section{Introduction}

The detection of gravitational waves (GW) opened a new window for observing the universe. Hundreds of signals have been detected by the laser interferometer gravitational wave detector network, including Advanced LIGO \cite{aasi2015advanced}, Advanced Virgo \cite{acernese2014advanced}, and KAGRA \cite{aso2013interferometer} during the first four data takings. Ground breaking scientific results spans from general relativity to astrophysics and cosmology. The scientific interest of detecting more and farther gravitational waves led to the proposals of next generation detectors such as Cosmic Explorer \cite{reitze2019cosmic}, Einstein Telescope \cite{punturo2010einstein} and Neutron Star Extreme Matter Observatory(NEMO) \cite{ackley2020neutron}. Both current and new-generation detectors will be mainly limited by quantum noise and the use of squeezed states of vacuum, injected from the dark port of the detector is a crucial technique for its reduction. 

Optical losses are extremely detrimental for the use of squeezed states of light \cite{kwee2014decoherence} and it has been shown that they represent the ultimate sensitivity limitation for a quantum-noise-limited detector \cite{miao2019quantum}.

Optical losses in Fabry-Perot cavities used in GW detectors are mainly due to imperfections and contamination of the mirrors' surface introduced during three main steps: polishing, coating, and integration of mirror in the experiment set-up. The characterization of mirror surface at each step is crucial for understanding the origin of such imperfections, and how they can be reduced. Ion-beam figuring, as a polishing technique, provides a deterministic method for achieving atom-level finishing of optical surfaces \cite{wilson1987neutral}, nevertheless sub-millimeter point defects were found to be brought by the coating and can strongly increase the optical losses \cite{brooks2021point}. 

The integration process can introduce contamination, which can be mitigated if proper cleaning countermeasures are taken \cite{gushwa2014coming}. For cryogenic GW detectors, optical losses caused by a uniform ice layer growing on the mirror surface was reported \cite{tanioka2021optical} as well as contamination from residual molecules in the vacuum system, which are attracted by mirrors at cryogenic temperature \cite{tanioka2020optical}. If mirrors are electrically charged they can attract such molecules even at room temperature \cite{Paoletti2023}. Moreover, it has been found that even laser can induce contamination \cite{D2017Laser}.

After the integration of mirrors in the cavity, \textit{in-situ} mapping of optical losses was done for several experiments, for instance 10\,cm-long \cite{cui2017simultaneous}, 80\,cm-long \cite{truong2019near} and 1.8\,m-long \cite{gutierrez2023optical} cavities. The dependence of optical losses on the beam position on the cavity mirror was observed for example in 300 meters \cite{capocasa2018measurement} and 38 meters optical cavities \cite{drori2022scattering}. However, a precise mapping of the optical losses as a function of the beam position on the mirrors was never done for a 100 m-scale cavity.

In this paper, we will report on the mapping of optical losses in the Virgo filter cavity \cite{acernese2023frequency}. Such a 285\,m suspended cavity is used in combination with a squeezed vacuum source to produce frequency dependent squeezing, a technique  which is able to reduce quantum noise in the whole bandwidth \cite{ganapathy2023broadband}. Given the large finesse ($\simeq 10000$) required for achieving optimal frequency dependency, cavity round trip losses (RTL) play a major role in the squeezing degradation and having a sound characterization and a good understanding of their origin is important to possible reduce them as much as possible in the future. 

The alignment strategy used for the control of such a cavity, allows to extract the intra-cavity beam position on the cavity mirrors in a precise and repeatable way. This allows us to perform a beam scan on input and end mirror respectively, while keeping the beam fixed on the other mirror. 

This article is organized as follows. In section \ref{sec:setup}, we introduce the measurement setup. In section \ref{sec:scan_method}, we present the methodology to locate the beam position on cavity mirrors and realize a beam scan.  In section \ref{sec:RTL_method}, the optical losses characterization method is described, while in \ref{sec:RTL_result} the results of such characterization are reported. In section \ref{sec:simulation}, we present the expected losses based on the characterization of mirrors properties before the integration in the cavity. Finally in section \ref{sec:discussion}, the results are discussed.

\begin{figure*}[t!]
    \includegraphics[width=1\textwidth]{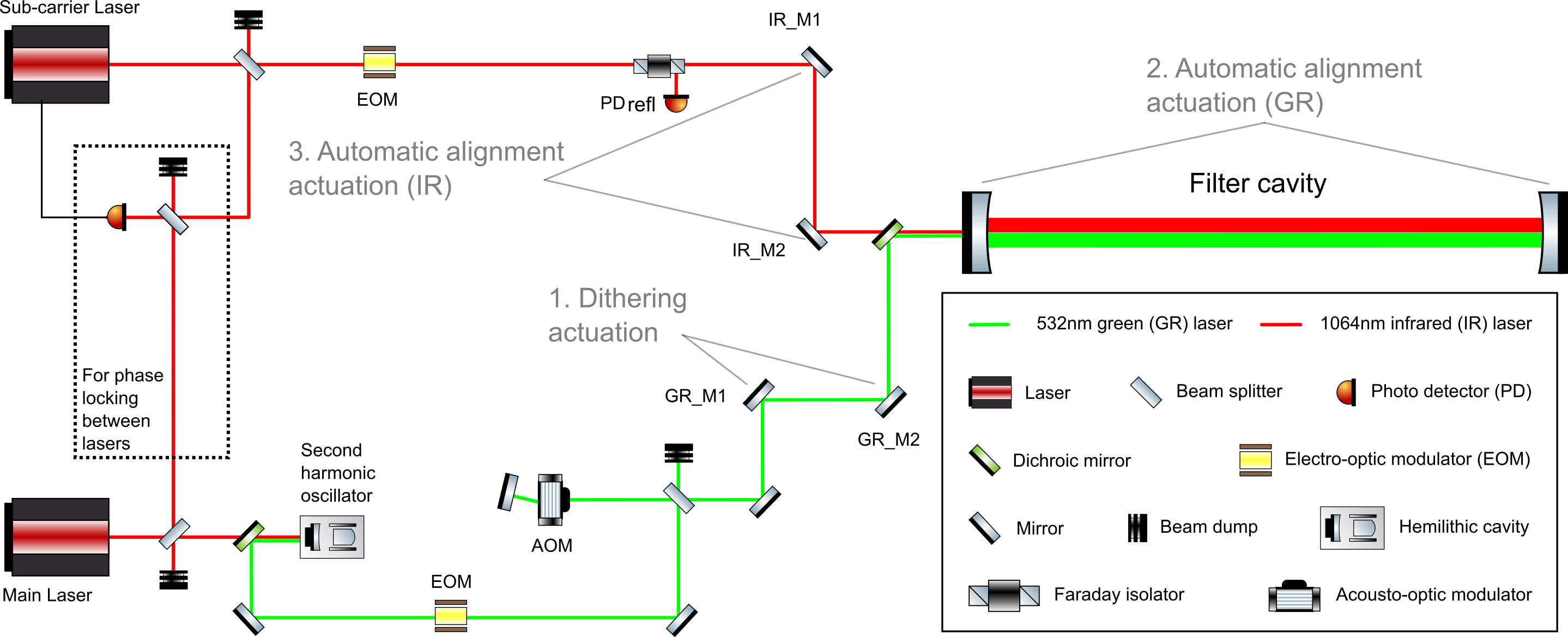}
    \caption{Experiment scheme. The beam scan consists of three steps. Firstly, the input beam direction is changed of the desired amount by tilting GR\_M1 and  GR\_M2 of the appropriate angle, producing a misalignment. Secondly the automatic alignment loop adjusts the filter cavity mirrors angle to realign the cavity axis with respect to the input beam. In this way, green beam arrives on the mirrors at the position where we want to measure optical losses. Thirdly, infrared automatic alignment moves IR\_M1 and IR\_M2 to make the IR input beam beam superpose to the new cavity axis. Finally the round trip losses are measured from the difference in the reflected power recorded by a photodiode in reflection of the cavity (named $\mathrm{PD_{refl}}$ in the scheme), when the IR beam is on-resonance and when it is off-resonance.
    }
\label{fig:scheme_experi}
\end{figure*}

\section{Experimental set-up}
\label{sec:setup}
A simplified experimental set-up is shown in Fig.\,\ref{fig:scheme_experi}, where only the components related to this work are shown. A detailed description of the full set-up, including the source of the squeezed vacuum beam which is reflected by the filter cavity before being injected into the interferometer, is reported in \cite{acernese2023frequency}.

The filter cavity is composed by two suspended mirrors, whose properties are summarized in Tab.\,\ref{tab:FCpara}. The squeezer main laser (wavelength 1064 nm, IR) is sent into a second harmonic generator, producing a green beam (GR) at 532 nm. A part of this green beam is used to pump the OPO for producing the squeezed vacuum beam while the other part is used for the angular and longitudinal control of the filter cavity. This beam goes through a double-passed acousto-optic modulator (AOM) \cite{donley2005double} to get a frequency shift, so that GR and IR beam can co-resonate inside the filter cavity. Thanks to the double pass scheme the GR field direction is insensitive to the AOM frequency variation. The GR beam, which has a finesse 100 times lower with respect to the IR, is kept on resonance into the filter cavity using the Pound Drever Hall (PDH) technique \cite{black2001introduction}, by making the cavity length follow the laser frequency. We remark that the use of GR beam is not mandatory for the IR losses characterization, but it makes it simpler to perform the measurements as the cavity can be easily kept locked.

An additional IR field (so-called sub-carrier) is generated from an auxiliary laser which is phase locked to the main laser with about 1.2\,GHz frequency offset. Such beam has been added to be used as an alternative to the GR beam for the cavity control and it is used in this work as a probe to measure the filter cavity round trip losses at 1064\,nm. It is kept on resonance with the cavity acting on the set point of its phase lock loop with the main laser.

\begin{table}[t!]
\caption{\label{tab:FCpara}Summary of the filter cavity parameters.}
\begin{ruledtabular}
\begin{tabular}{lcr}
Parameter & Symbol & Value \\
\hline
Length & L & \SI{284.9}{m}\\
Mirror diameter &  & \SI{149.9}{mm}\\
Input mirror radius of curvature & $R_1$ & \SI{556}{m}\\
End mirror radius of curvature & $R_2$ & \SI{557.1}{m} \\
\hline
\multicolumn{3}{c}{\textsc{Parameters for green}} \\
Input mirror transmissivity & $T_1$ & 2.6 \% \\
End mirror transmissivity & $T_2$ & 2.7 \% \\
Finesse & $\mathcal{F}$ & 117 \\
Beam radius at input mirror & $w_{1}$ & 7.4 mm \\
Beam radius at end mirror & $w_{2}$ & 7.4 mm \\
\hline
\multicolumn{3}{c}{\textsc{Parameters for infrared}} \\
Input mirror transmissivity & $T_1$ & 562 ppm \\
End mirror transmissivity & $T_2$ & 3.9 ppm \\
Finesse & $\mathcal{F}$ & 10020 \\
Linewidth & $\Delta_\nu$ & \SI{52.5}{Hz} \\
Storage time & $\tau$ & \SI{6.06}{ms} \\
Beam radius at input mirror & $w_{1}$ & 10.5 mm \\
Beam radius at end mirror & $w_{2}$ & 10.5 mm
\end{tabular}
\end{ruledtabular}
\end{table}

The alignment of the filter cavity includes three control loops: 

\begin{itemize}
    \item The first one use a wavefront sensing technique (WFS) \cite{morrison1994automatic} \cite{heinzel1999automatic}to keep the cavity axis aligned with the incoming GR beam axis, by acting on the cavity mirrors.
    \item The second one uses the so-called dithering technique \cite{yu2019astrophysical} to keep the incoming GR beam centered on the cavity mirrors by acting on the input beam direction through the steering mirrors GR\_M1 and GR\_M2. A detail description of this technique can be found in App.\,B. The calibrated error signals of this loop provides the information of the position of the beam on each of the mirrors.
     \item The third one uses again the wavefront sensing technique to keep the incoming IR beam superposed to the cavity axis (which is the same for IR and GR beam \footnote{This might be not exactly true as the mirrors are not perfectly spherical and the GR beam dimension is larger by a factor $\sqrt{2}$}, by acting on the steering mirrors IR\_M1 and IR\_M2)
\end{itemize}

The sensors used for such angular control loops are not shown in Fig.\ref{fig:scheme_experi} but they can be found in the complete scheme of the experiment in \cite{acernese2023frequency}.

\begin{figure*}[t!]
    \includegraphics[width=0.95\textwidth]{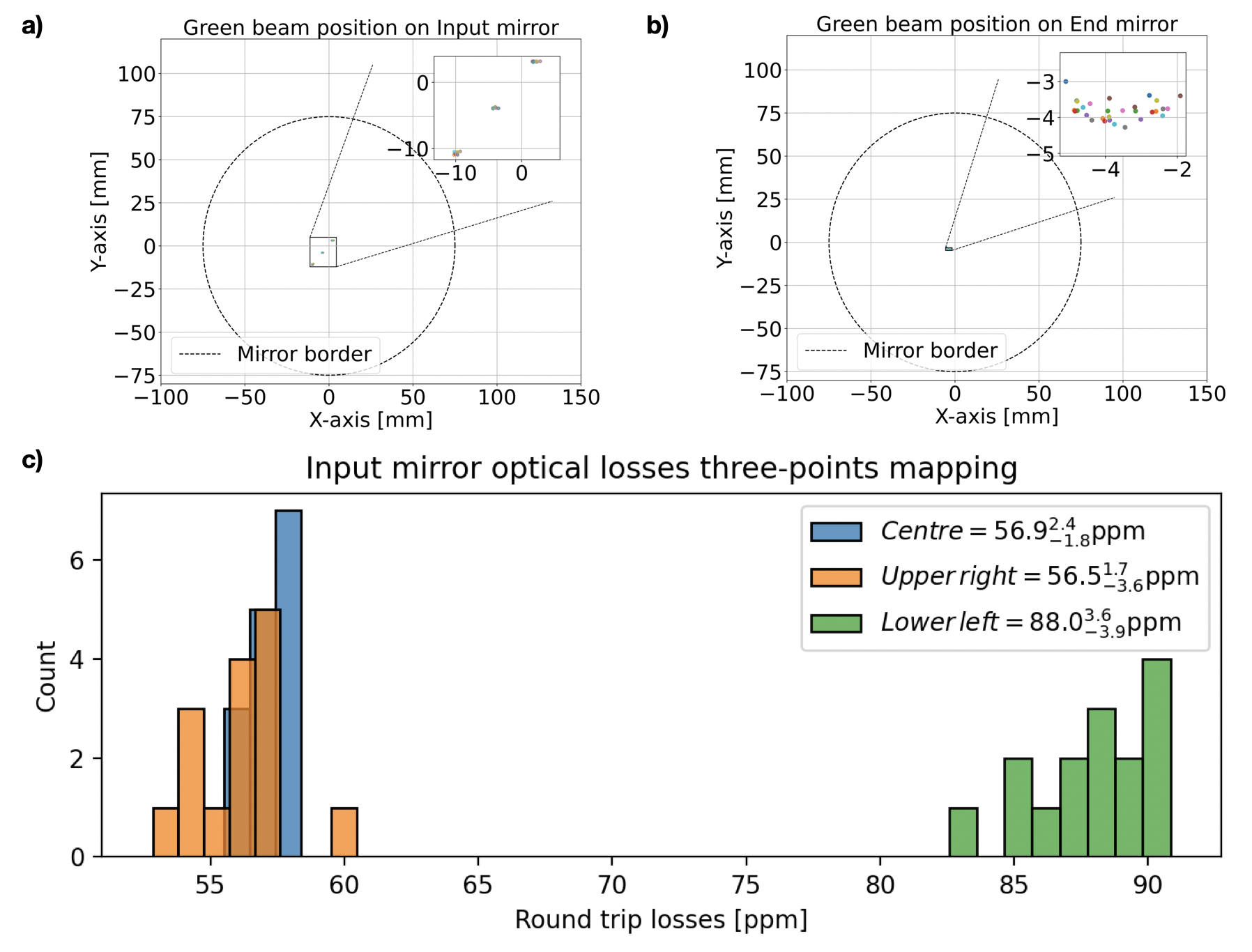}
    \caption{Round trip losses measurement for three points on the input mirror. \textbf{a)} Reconstruction of intra-cavity beam hitting position on the input mirror. \textbf{b)} Reconstruction of intra-cavity beam hitting position on the end mirror. \textbf{c)} Round trip losses for three points on the filter cavity input mirror.}
\label{fig:map_3}
\end{figure*}

\section{Beam scan implementation}
\label{sec:scan_method}

We measured a RTL map by performing a beam scan on input and end mirror respectively, while keeping the beam fixed on the other mirror. The fixed beam positions on the cavity mirrors are the middle points of Fig.\,\ref{fig:map_3} a) and b). These points correspond to the beam positions where we usually operate the cavity. During the scan the dithering loop was open and the incoming GR beam was opportunely moved using GR\_M1 and GR\_M2. The WFS GR alignment loop was active and it was moving the mirrors to keep the cavity axis aligned on the incoming beam. This results in a change of the beam position on the mirrors. The dithering error signal was used to know such position, as detailed in App.\,B. The steering mirror actuators have been properly calibrated in order to produce the desired change in the beam position on the mirrors. The relation between mirror angular motion and cavity axis tilt/shift depends on the cavity geometry and it is detailed in App\,A. A proper combination of actuators motions allow keeping the beam fixed on one mirror while moving it on the other one. The details of the implementation of the beam scan are given in App\,B. During the scan the IR beam was kept aligned with its WFS loop. Due to the limitation of the angular range of the GR steering mirror, the side of the mapping area that we could scan is about 2\,cm. 

\section{Round trip losses measurement method}
\label{sec:RTL_method}
For each point of the scan the RTL have been extracted from the comparison of the reflectivity of the cavity when IR is on-resonance and and when it is off resonance\cite{capocasa2018measurement}. Although the end mirror is almost completely reflective ($T_2$ = 3.9\,ppm, which will be included in the RTL), there is a reduction of reflected power when the cavity goes from off-resonance to on-resonance. This power reduction is only due to the RTL, and  we can extract their value information using the formula

\begin{equation}
    L \simeq \frac{\rm{T_1}}{2}\frac{1-\rm{R_{cav}}}{1+\rm{R_{cav}}},
\end{equation}

where $\rm{R_{cav}}=\rm{\frac{P_{on}}{P_{off}}}$. $\rm{P_{on}}$ is the power detected on $\rm{PD_{refl}}$ (see Fig.\ref{fig:scheme_experi}) when the cavity is on-resonance, while $\rm{P_{off}}$ is the power detected on the same photo detector when the cavity is off resonance. For each measurement we did several switches between on and off resonant conditions. For each stretch of data we computed a statistical error which was found to be always below 1.4\,ppm.

The systematic error for this on-off resonance method comes from the uncertainty of two quantities: input mirror transmissivity and the amount of light which doesn't resonate in the cavity (called uncoupled power). The input mirror transmissivity has been measured after coating at the Laboratoire des Matériaux Avancés (LMA) in Lyon, finding a value of 562$\pm$1\,ppm \cite{input}. The uncoupled power (mainly coming from a mismatch between the IR beam mode and the filter cavity mode and the presence of modulation sidebands)can be easily estimated by measuring the power in the higher order modes and in the non-resonant sidebands when the laser frequency is scanned over one free spectral range. We have measured the uncoupled power to be 2$\pm$1\%. These two sources of systematic error are taken into account in the RTL calculation (using the Python package \cite{eric2022}), obtaining a systematic error of 1.3\,ppm. A detailed explanation of how the uncoupled power affects the RTL estimation can be found in Sec IV for Ref. \cite{capocasa2018measurement}

\section{RTL measurement results}
\label{sec:RTL_result}
In order to make a first check of the repeatability of our procedure we perform several RTL measurements at 3 different points on the input mirror. The three points are spaced by 9.3\,mm with respect to each other in a straight line. For all the measurements done, we plot the beam position measured with the GR dithering signal on the input and on the end mirror, as shown in Fig.\,\ref{fig:map_3} \textbf{a)} and \textbf{b)} respectively. RTL measurements were performed a total of 10 times in different moments over a period of 10 days for each of the 3 points and the result are shown in \textbf{c} of Fig.\,\ref{fig:map_3}. This measurement shows that there is a clear and reproducible dependence of the RTL level with respect to the beam position. Note that the Fig.\,\ref{fig:map_3} \textbf{b)} shows that the beam position reconstruction on the filter cavity end mirror spreads out in the horizontal direction. This is likely due the hysteresis of piezo actuators used to move the GR steering mirrors.

\begin{figure*}[t!]
    \includegraphics[width=0.95\textwidth]{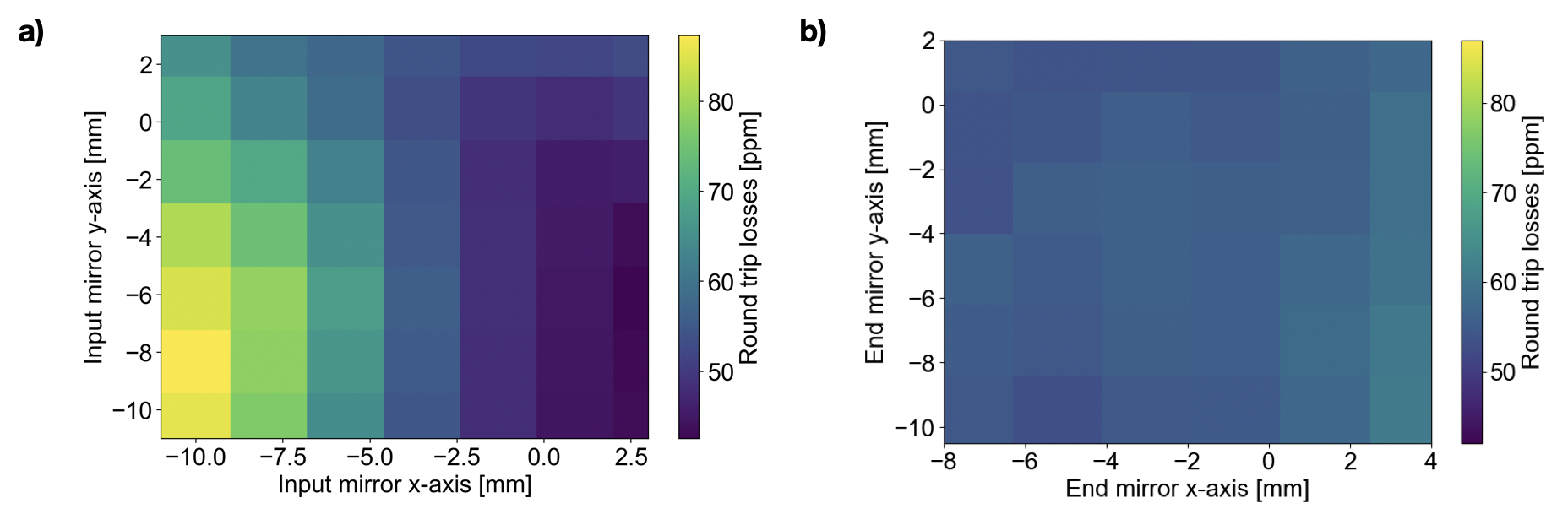}
    \caption{Measured round trip loss maps obtained changing the beam position on the input mirror (left plot) and on the end mirror (right plot).}
    \label{fig:map_map}
\end{figure*}

After this, we performed a RTL map, scanning the beam position of the zoomed-in area in Fig.\,\ref{fig:map_3} \textbf{a)} with a step size of 2.2\,mm, both on input and end mirror respectively. We chose a relatively small value of the step size to locate precisely the potential contamination sources and to find exactly where RTL is minimum. The measured RTL maps are shown in Fig.\,\ref{fig:map_map}. The RTL on input mirrors ranges from 42\,ppm to \,87 ppm while for the end mirror ranges from 53\,ppm to 61\,ppm. For realizing each map, the beam has been kept fixed at the usual operation point
on the other mirror. We remark that a significant spatial dependence of optical losses was only found on the input mirror.

\section{Estimation of the scattering losses from mirrors' characterization}
\label{sec:simulation}

\begin{figure*}[t!]
    \includegraphics[width=0.7\textwidth]{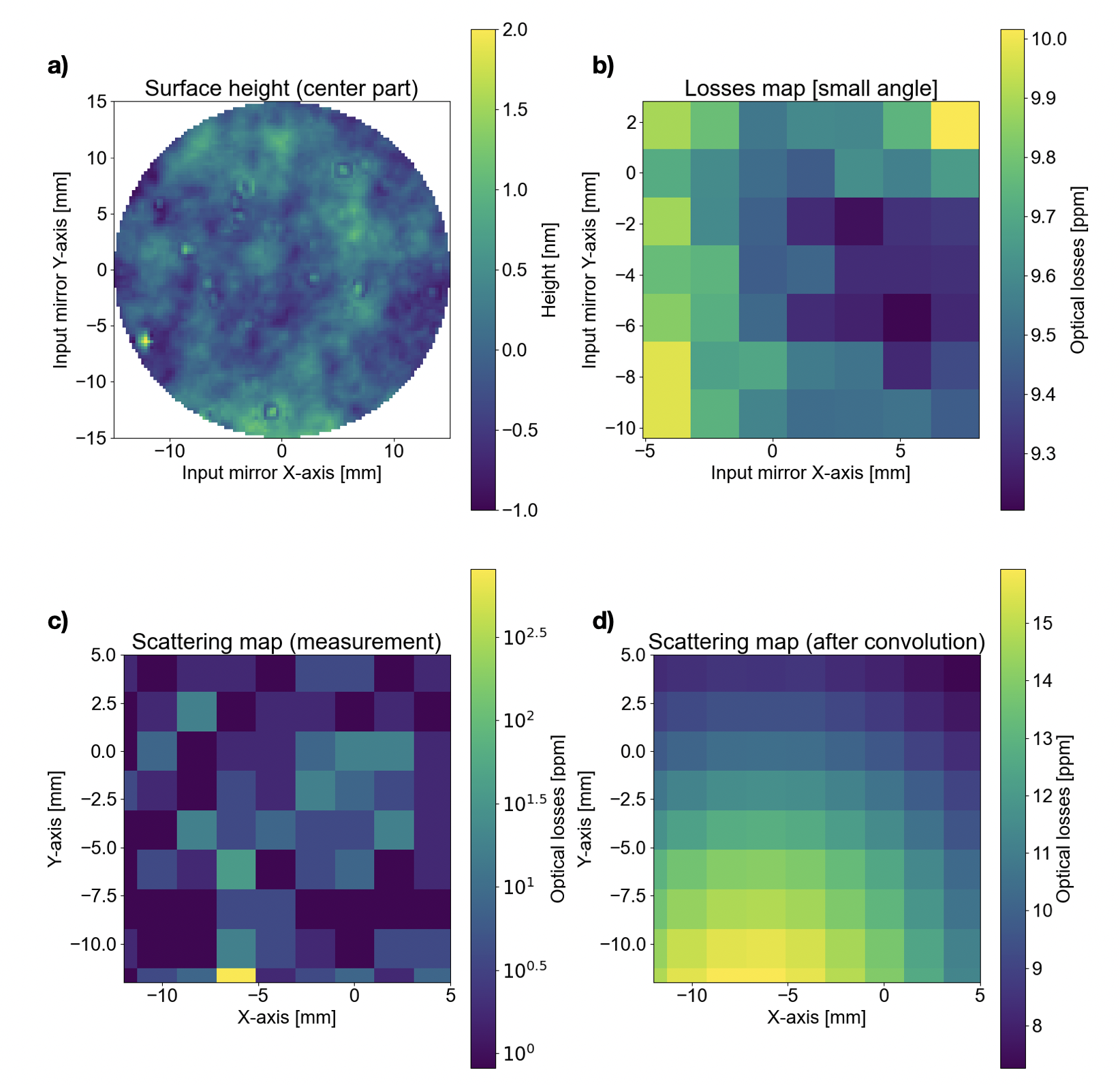}
    \caption{Analysis of small-angle and large-angle scattering. \textbf{a)} Surface map of the input mirror, measured after polishing, \textbf{b)} OSCAR simulation of small-angle scattering using the surface map on the left and, \textbf{c)} Measurement of large angle scattering for input mirror using scatterometer. Note that the colorbar here is in log scale. The shown region is consistent with the region where optical losses are measured in the Virgo filter cavity, as shown in the in-set in Fig.\,\ref{fig:map_3} \textbf{a)}. \textbf{d)} Large angle scattering map after making a convolution of Gaussian beam intensity function with the scatterometer result.}
    \label{fig:estimation}
\end{figure*}

An estimation of the expected filter cavity losses can be done relying on the mirrors' characterization realized before the cavity integration. As described in \cite{capocasa2016estimation}, a Zygo interferometer is used to check the shape deviation of the mirror with respect to a perfect sphere after mirror polishing. The recorded map, often referred to as a \textit{flatness} or \textit{waviness} map, is shown in Fig.\,\ref{fig:estimation}\,a), and has a spatial resolution $\Delta$ of 0.38\,mm. Consequently, the maximum spatial frequency $\mathrm{f_{max-map}}$ reachable from this map is defined by
\begin{equation}
    f_{max-map} = \frac{\mathrm{N}}{2}\Delta \mathrm{f} = \frac{\mathrm{N}}{2}\frac{1}{\mathrm{N}\Delta} = \frac{1}{2\Delta}
\end{equation}
i.\,e.\,1300\,m$^{-1}$. This spatial frequency is related to the scattering angle by the general relation
\begin{equation}
  f = \frac{\sin{\theta}}{\lambda}.
\end{equation}
which means that the scattering angles associated to these shape errors are less than 0.1$^{\circ}$ (small angle scattering). 

Such a \textit{waviness} map, which is measured after polishing, is consider to be valid also after the coating process. This is because the ion-beam sputtering technique guarantees that the surface flatness is reproduced when coating is applied \cite{straniero2015realistic}\footnote{with the exception of a change in the overall radius of curvature (for example the Virgo filter cavity mirrors are found to be 0.34\% more curved after coating)}. We used such map in the FFT optical simulation code Oscar \cite{OSCARx} to estimate the small angle scattering losses. In the simulation, we propagate a beam into a cavity with the same parameters of the Virgo filter cavity and we calculate the power losses of the beam due to the multiple reflection by imperfect mirrors. By centering the beam on different positions on the input mirror flatness map in Oscar, we obtain the RTL map shown in Fig.\,\ref{fig:estimation}. The scanned region is identical to the experimental measured region shown in Fig.\,\ref{fig:map_3} a). In this simulation the losses vary from 9.2 to 10.0\,ppm. 

In order to characterize the large-angle scattering, which can be increased by the presence of point defects introduced during the coating process, a raster scan of the scattered light is realized at LMA after the coating is applied, with a dedicated device (the CASI scatterometer). Such instrument uses a characterization laser with small beam size, so that the resolution is 2\,mm. Such map is realised as follows:

\begin{itemize}
    \item The light scattered by a small area of the mirror at $\theta_0$ = 10$^{\circ}$, $\phi_0$ = 0$^{\circ}$ with respect to the specular beam is measured using a CASI scatterometer with a spot size of 2\,mm in diameter. The incident beam is scanned at the surface of the mirror to provide a map ARS(x,\,y;\,$\theta_0$,\,$\phi_0$) of the angle-resolved value for that particular direction in the plane of incidence.
    \item For a particular point (x$_0$,\,y$_0$) of this map, that has a value of ARS(x,\,y;\,$\theta_0$,\,$\phi_0$) close to the average one, the angle dependence of ARS is recorded, always in the plane of incidence ($\phi_\mathrm{d}$=0), but between $\theta_\mathrm{min}$ and 90$^{\circ}$. We assume that the angle-resolved scattering of the mirror surface has no dependence on the azimuthal angle ($\phi_\mathrm{d}$), which is true if this scattering is induced by the roughness of micropolished optics. Thus, we can derive the total integrated scattering of this specific point using the following formula
    \begin{eqnarray}
        && \mathrm{TIS}(\mathrm{x_0},\mathrm{y_0}) \nonumber \\
         &=& \int_{0}^{2\pi}\int_{\theta_{\mathrm{min}}}^{\pi/2}\mathrm{ARS}(\mathrm{x_0},\,\mathrm{y_0};\,\theta,\,\phi)\sin{\theta}\mathrm{d}\theta \mathrm{d}\phi \nonumber \\
        &=& 2\pi \int_{\theta_{\mathrm{min}}}^{\pi/2}\mathrm{ARS}(\mathrm{x_0},\,\mathrm{y_0};\,\theta,\,\phi_0)\sin{\theta}\mathrm{d}\theta
    \end{eqnarray}
    \item Recording the ARS dependence on the scattering angle $\theta_\mathrm{d}$ for all points of the map is impossible because it is too time-consuming. Therefore, our approach is to assume that the ARS has the same angular dependence for each point on the mirror and to use the TIS value obtained from this representative point to derive the TIS value for all the points on the raster scan map, using the following formula
    \begin{eqnarray}
        \mathrm{TIS}(\mathrm{x},\mathrm{y}) = \mathrm{TIS}(\mathrm{x_0},\mathrm{y_0})\frac{\mathrm{ARS}(\mathrm{x},\,\mathrm{y};\,\theta_0,\,\phi_0)}{\mathrm{ARS}(\mathrm{x_0},\,\mathrm{y_0};\,\theta_0,\,\phi_0)}
    \end{eqnarray}
\end{itemize}

It is clear that the assumption that the BRDF has the same angular dependence for all the analyzed points on the surface of the coated mirror is problematic, since micro-roughness, point defects and contaminants have different scattering distributions, but it is the only way available to us to evaluate this TIS map. For a more accurate and realistic determination, it would be necessary to use either an integrating sphere measurement with scanning capability\,\cite{schroder2005measurement}, or even better a scatterometer with imaging ability, such as SPARSE \cite{lequime2009goniometric}\cite{bolliand2022new}.

In the Virgo filter cavity, the beam radius of the resonating IR Gaussian beam is $10.5\,$\,mm. Thus, in order to have a more realistic estimation of the large-angle scattered light after the integration of mirrors in the filter cavity, we made a convolution of the scatterometer results with a Gaussian intensity function as the one of the intra-cavity beam. The convolution has been done on an area of 17$\times$ 17\,mm, which corresponds to the mirror surface for which RTL maps has been measured, as described before. The result is shown in c) and d) of Fig.\,\ref{fig:estimation}. The large angle scattering map before convolution ranges from 1\,ppm to 884\,ppm. The estimated RTL due to large angle scattering are between 7\,ppm to 16\,ppm for input mirror and 8\,ppm for end mirror, giving a total of 14-24\,ppm . Due to the device limitation, the angle scattering measured by CASI is above $\simeq 3\, $ deg. This corresponds to a spatial frequency $\mathrm{f}_{\mathrm{min-CASI}} = 49188\,m^{-1}$. A device that could resolve the ARS both spatially and angularly, such as mentioned above \cite{lequime2009goniometric}\cite{bolliand2022new} would greatly improve the characterization accuracy. However, the current large angle scattering map serves as a preliminary estimate for this work.

\begin{figure}[t!]
    \includegraphics[width=0.5\textwidth]{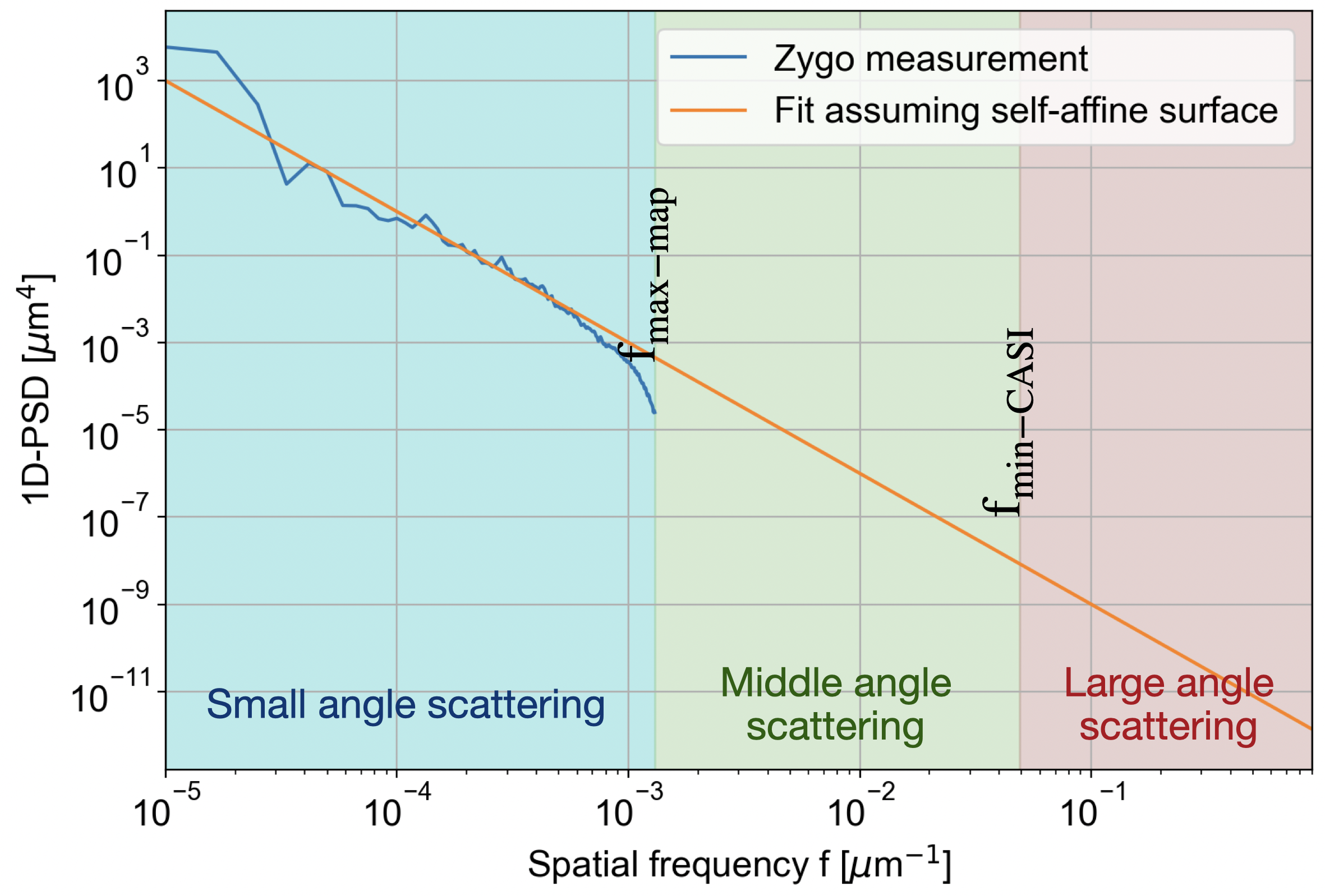}
    \caption{One dimension PSD from the mirror surface height measurement and a fit of it. The whole figure is separated into three regions: small, middle, large angle scattering regions. Some important frequencies are mentioned as well.}
\label{fig:psd}
\end{figure}

From the discussion above, we remark that the scattering losses between 0.1 degrees and 3 degrees (we will call them "middle angle scattering") are missing in our estimation. This corresponds to the green shaded area in Fig.\, \ref{fig:psd}. To infer this information, we adopt the following method: we extrapolate the mirror surface power spectral density (PSD) at the frequencies corresponding to this angle region, we use scattering model to estimate the corresponding losses. The extrapolation is based on the flatness maps measured before coating is applied using a power law, which is plotted in Fig.\,\ref{fig:psd}.

The one dimensional power spectral density was obtained with the following procedure
\begin{itemize}
    \item Use Oscar software to remove the offset, radius of curvature, and tilt from the flatness mirror map
    \item Apply Hann window to Oscar processed surface and perform two dimensional Fast Fourier Transform for windowed surface to obtain the spectral density 
    \item Calculate the radial average of the two dimensional PSD assuming the surface is radially symmetric and get one dimensional PSD 
\end{itemize}

It is now well established that micro-polished optics are self-affine \cite{jacobs2017quantitative} in the spatial frequency range between the inverse of the size of the optics and the inverse of the wavelength. This allows us to describe the spatial frequency dependence of the 1D-PSD of the mirror surface as a straight line in a log-log representation (see Fig.\,\ref{fig:psd}).

From the integration of power spectral density between $\mathrm{f}_{\mathrm{max-map}} = 1300 \mathrm{m}^{-1} $ and $\mathrm{f}_{\mathrm{min-CASI}} = 49188 \mathrm{m}^{-1}$ which corresponds to the middle angle scattering frequencies we found $\sigma_{\mathrm{middle}}$, the RMS deviation of the surface from a perfect spherical shape to be:

\begin{equation}
\label{eqn:sigma}
    \sigma_{\mathrm{middle}}^2 = 2\pi\int_{\mathrm{f}_{\mathrm{max-map}}}^{\mathrm{f}_{\mathrm{min-CASI}}}\mathrm{PSD(f)f} \,df = 0.069\,\mathrm{nm}
\end{equation}

which can be used in the following formula to estimate the middle angle scattering losses\footnote{We can use this formula as the The Rayleigh smooth-surface criterion \cite{stover1995optical}is verified. It requires $\frac{4\pi\sigma\cos{\theta_{i}}}{\lambda}^2 \ll 1$. In our case the incident angle $\theta_i$ = 0, so that the value is about $10^{-5}$.}:

\begin{equation}
\label{eqn:tis}
    \mathrm{TIS_{\mathrm{middle}}} = \left(\frac{4\pi\sigma_{\mathrm{middle}}}{\lambda}\right)^2 \simeq 0.7\,\mathrm{ppm} 
\end{equation}

which become 1.4 ppm, when considering the contribution of both mirrors. The contribution of the middle angles scattering is found to be quite small but this estimation does not include the effect of any defect that might be introduced during the coating application. 

In order to obtain the total estimation we add 3.9\,ppm of transmission from the end mirror and we neglect the absorption and clipping losses which are expected to be below 1\,ppm. We predict a total round trip losses between 30.3 and 39.3\,ppm. The detailed budget is summarized in the Tab.\,\ref{tab:sigma_loss}: 

\begin{table}[h!]
\caption{\label{tab:sigma_loss} Estimated losses from mirrors characterization before cavity integration}
\begin{ruledtabular}
\begin{tabular}{lr}

Small angle scattering &  10 ppm \\
Middle angle scattering &  1.4 ppm\\
Large angle scattering &  15-24 ppm \\
End mirror transmission &  3.9 ppm\\
Absorption and clipping  &  $<1$ ppm\\
\hline
\textbf{Total}  &  \textbf{30.3-39.3 ppm}
\end{tabular}
\end{ruledtabular}
\end{table}

\section{Conclusions}
\label{sec:discussion}


We have characterized the dependence of cavity round trip losses on the beam position on mirrors ranging from 42\,ppm to 87\,ppm with an estimation error of 4\,ppm. A clear dependence of optical losses on the beam position on the cavity mirrors has been found. 

The difference between the smallest losses measured during the scan of the beam position on the mirrors (42\,ppm) and the losses expected from the mirrors characterization prior to the cavity assembly (30-39 ppm) is quite small suggesting that there is no major systematic effect that has not been taken into account in the estimation. We believe that the increased losses for some beam positions on the input mirrors might be due to some contamination introduced during the cavity integration process. 

In the future, achieving low optical losses is crucial for the performance of gravitational wave detectors. We think that the \textit{in-situ} measurement method presented in this paper is not only useful for a thorough cavity loss characterization but also for the actual reduction of the losses, thanks to the optimization of the beam position. Moreover this losses characterization is crucial to localize possible contamination particles and verifying the effectiveness of \textit{in-situ} cleaning techniques, such as gas jet to be used within the vacuum system \cite{gasjet}.

\FloatBarrier
\section*{Acknowledgements}
We acknowledge the Quantum Noise Reduction (QNR) team of Advanced Virgo for the realisation the Frequency Dependent Squeezing system in Advanced Virgo, and in particular, J.P. Zendri, M. De Laurentis, F. Sorrentino, R. Bonnand, A. Bertolini, M. Tacca for having managed the activity, H. Vahlbruch and M. Mehmet for having implemented the squeezing source and the staff of the European Gravitational Observatory (EGO) for the contribution to the logistic and the system integration. We thank L.Pinard, C.Michel and the Laboratoire des Materiaux Avancés (LMA) for the mirrors realisation and characterisation and Degallaix for the fruitful discussions and the useful comments on the manuscript. This work has been supported by ANR-23-CE31-0004 and by the EU Horizon 2020 Research and Innovation Program under the Marie Sklodowska-Curie Grant Agreement No. 101003460.

\appendix

\section{Relation between cavity axis direction and beam position on the mirrors }

The cavity axis, defined as the line passing by the two center of curvature of the mirrors, determines the direction of the resonant beam and thus the hitting point of the beam on the mirrors. In order to make a scan of the beam on the mirrors we need to opportunely change the cavity axis direction by changing input mirror tilt angles ($\alpha_1$) and end mirror tilt angles ($\alpha_2$).\footnote{Since the effect  of mirror tilts around the two axes orthogonal to the beam direction will have identical expressions we consider here only one case} The cavity axis direction is usually described as a combination of tilt ($\alpha$) and shift ($d$) of the axis itself,  with respect to the position in which the axis passes through the physical center of each mirror. Mirrors' tilts and cavity axes tilt and shifts are related by the cavity geometry \cite{bassan2014advanced} 

\begin{equation}
\label{eqn:mir_cav}
    \smallMatrix{\alpha \\ d}  = \smallMatrix{ \frac{-R_1}{L-R_1-R_2} & \frac{R_2}{L-R_1-R_2} \\ \frac{R_1(R_2-L)}{R_1+R_2-2L} & \frac{R_2(R_1-L)}{R_1+R_2-2L} } \smallMatrix{\alpha_1 \\ \alpha_2}
\end{equation}

\begin{figure}[t!]
    \includegraphics[width=0.45\textwidth]{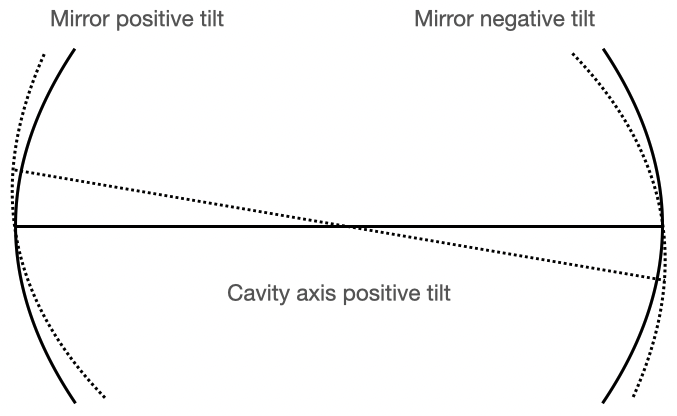}
    \caption{Direction convention when cavity alignment is changed. The solid lines represents the  situation when the cavity axis passes by the center of the mirror. When the mirror are tilted the cavity axis direction changes, as represented by the dotted lines.}
\label{fig:convention}
\end{figure}

In the Virgo filter cavity, input mirror radius of curvature $R_1$ is 556\,m, the end mirror radius of curvature $R_2$ is 557.1\,m, and cavity length $L$ is 284.9\,m. Thus, we get the following relations for our cavity
\begin{eqnarray}
    \alpha &=& 0.671\alpha_1 - 0.673\alpha_2 \nonumber \\
    d &=& 278.6\alpha_1 + 278.0\alpha_2
\end{eqnarray}
The above equation can be approximated as
\begin{eqnarray}
    \alpha &=& 0.67 (\alpha_1 - \alpha_2) \nonumber \\
    d &=& 278 (\alpha_1 + \alpha_2)
\end{eqnarray}
Tilting both mirrors of the same amount in the same direction will produce a pure cavity axis shift while tilting both mirrors of the same amount in opposite directions will produce a pure cavity axis tilt. The convention of mirror tilt sign is shown in Fig.\,\ref{fig:convention}.

To convert mirror tilts to beam position change on cavity input and end mirrors, named $\Delta^{i}$ and $\Delta^{e}$ respectively, we use the following equations (obtained using small angle approximation): 
\begin{eqnarray}
\label{eqn:cav_dis}
    \Delta^{i} &=& \alpha z_0 + d \nonumber \\
    \Delta^{e} &=& -\alpha z_1 + d
\end{eqnarray}

where $z_0 = 142.8 $ m  and $z_1 = L- z_0 $ are the distance from filter cavity input mirror to the beam waist.  
Combining Eq.\,\ref{eqn:mir_cav} and Eq.\,\ref{eqn:cav_dis}, we obtain an expression to connect the mirror tilts to beam position change:

\begin{equation}
    \smallMatrix{\Delta^{i} \\ \Delta^{e}}  = \smallMatrix{z_0 & 1 \\ -z_1 & 1} \smallMatrix{ \frac{-R_1}{L-R_1-R_2} & \frac{R_2}{L-R_1-R_2} \\ \frac{R_1(R_2-L)}{R_1+R_2-2L} & \frac{R_2(R_1-L)}{R_1+R_2-2L} } \smallMatrix{\alpha_1 \\ \alpha_2}
\end{equation}
after inserting the numerical values, we obtain 

\begin{equation}
\label{eqn:mir_dis}
    \smallMatrix{\Delta^{i} \\ \Delta^{e}}  = \smallMatrix{374.4 & 182.0 \\ 183.1 & 373.6}\smallMatrix{\alpha_1 \\ \alpha_2}
\end{equation}
Here the unit of $\alpha_1$ and $\alpha_2$ are $\mu$rad, while the unit for $\Delta^{i}$ and $\Delta^{e}$ are mm.

From Eq.\,\ref{eqn:mir_dis}, we can easily infer that to keep the beam fixed on one of the cavity mirrors, we need to satisfy two conditions: the two mirrors tilts have to have opposite sign and one needs to be about twice as large as the other.

\section{Reconstruction of the beam position on the mirror using dithering technique}

Optical levers \cite{walet2022advanced} provide information about the change of mirrors' angular position, which can be used to reconstruct the beam position variation on the cavity mirrors. However, they don't provide the information about the absolute position of the beam on the mirror. Such information can be extracted using the fact that if the beam does not hit the mirror rotation center any angular movement of the mirrors will modify the effective length of the cavity and will be visible in the longitudinal control loop error and correction signals.

In practice, we modulate the angular position of the mirrors at a specified frequency and demodulate the locking correction signal at the same frequency. The amplitude of the demodulated signal is proportional to the beam's displacement from the center, reaching its minimum when the beam is hitting the center of the mirror. This sensing technique is usually referred to as \textit{dithering}.

\begin{figure*}[t!]
    \includegraphics[width=0.75\textwidth]{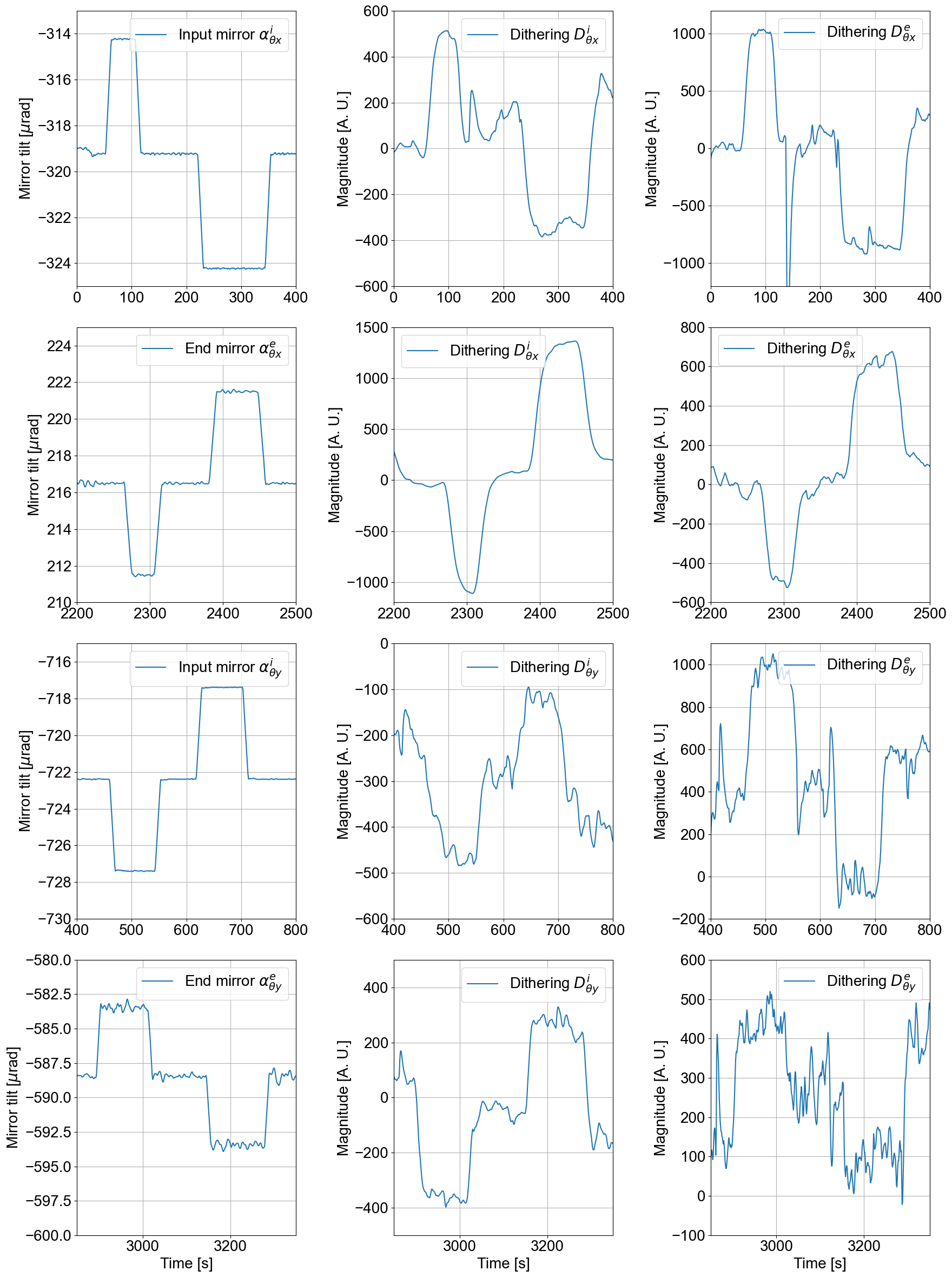}
    \caption{Measurements of calibration matrix. Each row corresponds to a degree of freedom: the left subplot shows the mirror tilt introduced on purpose, while the middle and right subplots show the induced dithering error signal variations.}
\label{fig:cal}
\end{figure*}

Such error signals have been calibrated relying on the calibration of the mirror angular actuators and on the information of the cavity geometry in Eq.\ref{eqn:mir_dis}. In order to perform such calibration a mirror tilt of 5\,$\mu$rad is introduced once a time around the $x$ and $y$ axis for the input and end mirror, as shown on the left column of Fig.\,\ref{fig:cal}. Each mirror tilt determines a change in two demodulated signals, as shown in the central and right column in Fig.\,\ref{fig:cal}. 
Such measurement allows to fill the matrix reported below, obtaining a calibration from mirror tilt $\alpha^{i,e}_{\theta x,\theta y}$ (unit $\mu$rad) to dithering signals $D^{i,e}_{\theta x,\theta y}$ (arbitrary unit)\footnote{Here the superscript $i$ and $e$ refer to the input and to the end mirror respectively while the subscript $\theta_x$ and $\theta_y$ refer to the angular positron change around the $x$ or the $y$ axis of the mirrors}. 

\begin{equation}
\label{eqn:matrix_t_s}
    \smallMatrix{D^{i}_{\theta x} \\ D^{e}_{\theta x} \\ D^{i}_{\theta y} \\ D^{e}_{\theta y}} =
    \smallMatrix{105.8 & 244.2 & 0 & 0 \\ 205.6 & 115.1 & 0 & 0 \\ 0 & 0 & 35.0 & -77.7 \\ 0 & 0 & -115.7 & 55.1} \smallMatrix{\alpha^{i}_{\theta x} \\ \alpha^{e}_{\theta x} \\ \alpha^{i}_{\theta y} \\ \alpha^{e}_{\theta y}}
\end{equation}
Since the alignment control scheme is such that the cavity axis is kept aligned on the input beam, a change in the input beam direction results in a change of the beam position on the mirror and thus in the dithering signals. For this reason it also useful to measure the transfer matrix between the error signal of the dithering control loop and the actuation voltage on GR\_M1 and GR\_M2.

\begin{equation}
\label{eqn:matrix_t_s}
    \smallMatrix{M^x_{1} \\ M^x_{2} \\ M^y_{1} \\ M^y_{2}} =
    \smallMatrix{0 & 0 & 1.72 & 1.17 \\ 0 & 0 & -0.47 & 1.05 \\ -0.74 & 0.76 & 0 & 0 \\ -0.19 & -0.64 & 0 & 0}    \smallMatrix{D^{i}_{\theta x} \\ D^{e}_{\theta x} \\ D^{i}_{\theta y} \\ D^{e}_{\theta y}} 
\end{equation}

\begin{table}[h]
\begin{ruledtabular}
\begin{tabular}{lcr}
beam position change & GR\_M1\,[V] & GR\_M2\,[V] \\
\hline
up on input mirror & 1.06 & 0.36 \\
up on end mirror   & -1& 0.73\\
down on input mirror & -1.06 & -0.36 \\
down on end mirror & 1 & -0.73 \\
left on input mirror & 0.85 &  -0.2\\
left on end mirror  & -0.87 & -0.72 \\
right on input mirror & -0.85& 0.2 \\
right on end mirror  & -0.87 & -0.72 \\
\end{tabular}
\caption{\label{tab:voltage_scan} Voltage required on GR steering mirrors M1 and M2 (unit:V) to perform a step of size 2.2\,mm on one cavity mirror, keeping the beam fixed on the other one.}
\end{ruledtabular}
\end{table}

According to these two matrices, we can see for example that in order to perform a step of 2.2\,mm, the voltages to be send to GR steering mirrors are those reported in table \ref{tab:voltage_scan}. 
Due to the fact the maximum voltage we can send to the steering mirrors is $\pm 9$\,V, the mapping area is limited to be less than 2\,cm. As shown in Fig.\ref{fig:cal}, demodulated dithering signals are affected by fluctuations which limit the precision at which we can measure the beam position. This error is measured to be $\pm$0.6\,$\mu$rad at maximum for each degree of freedoms. Accordingly, an error of $\pm$0.3\,mm in the beam position both on input and end mirror is assumed.The fact that we observed a spread in the reconstructed horizontal position of the beam on the end mirrors made us consider systematic errors as well. The main source of this systematic error should be the hysteresis of the piezo actuator that we use to move the GR steering mirrors.
The scan of the input/end mirror involves approximately RTL 300 measurements and it has been automatized using a python algorithm based on an environment (called PyALP) embedded in the Virgo data acquisition system. It takes about 30 minutes to perform the scan. The speed limitation comes from the fact that the angular control bandwidth of the cavity is less than 1\,Hz and it takes time to reach each new working point.  

\FloatBarrier
\bibliography{align_rtl}

\end{document}